\documentclass{Interspeech2024}

\usepackage{amssymb}
\usepackage{multirow}
\usepackage{multicol}
\usepackage{caption} 
\usepackage{algorithm}
\usepackage{algorithmicx}
\usepackage{dirtytalk}
\usepackage{algpseudocode}
\usepackage{graphicx} 
\usepackage{amsfonts}
\usepackage{caption}
\usepackage{subcaption}
\usepackage{censor} 
\usepackage{bm}
\usepackage{booktabs}
\usepackage{xcolor}
\usepackage{pifont}% http://ctan.org/pkg/pifont
\usepackage{xpatch}
\usepackage{comment}
\usepackage{tabularx}
\usepackage{enumitem}
\usepackage{censor}
\usepackage{threeparttable}

\usepackage{xstring}
\usepackage{collcell}
\usepackage{tikz}
\usepackage{textcomp}
\usepackage{xpatch}
\usepackage{pifont} % Required for \ding command

\newcommand{\xmark}{\ding{55}} % X mark

\def\B{{\cal B}}
\def\T{{\cal T}}

\interspeechcameraready

\title{Unsupervised Online Continual Learning for Automatic Speech Recognition}

\name{Steven}{Vander Eeckt}
\name{Hugo}{Van hamme}

\address{
    KU Leuven \\
  Department Electrical Engineering ESAT-PSI, Leuven, Belgium}
\email{\{steven.vandereeckt, hugo.vanhamme\}@esat.kuleuven.be}


\begin{document}

\maketitle

% the abstract here must exactly match the abstract entered into the paper submission system
\begin{abstract}
Adapting Automatic Speech Recognition (ASR) models to new domains leads to Catastrophic Forgetting (CF) of previously learned information. This paper addresses CF in the challenging context of Online Continual Learning (OCL), with tasks presented as a continuous data stream with unknown boundaries. We extend OCL for ASR into the unsupervised realm, by leveraging self-training (ST) to facilitate unsupervised adaptation, enabling models to adapt continually without label dependency and without forgetting previous knowledge. Through comparative analysis of various OCL and ST methods across two domain adaptation experiments, we show that UOCL suffers from significantly less forgetting compared to supervised OCL, allowing UOCL methods to approach the performance levels of supervised OCL. Our proposed UOCL extensions further boosts UOCL's efficacy. Our findings represent a significant step towards continually adaptable ASR systems, capable of leveraging unlabeled data across diverse domains.
\end{abstract}

\section{Introduction}

Artificial Neural Networks (ANN) have propelled Automatic Speech Recognition (ASR) to new heights. Despite their advancements, ANNs are hindered by Catastrophic Forgetting (CF) \cite{catastrophicforgetting}, which severely limits the capacity of ASR models to learn continually. Fortunately, Continual Learning (CL) has emerged as a solution to overcome CF in ANNs, showing considerable progress recently. CL enables models to learn from all accessible data, thus building powerful models that perform robustly across various accents, dialects, speakers, and domains in ASR, without requiring the reintroduction of all past data to prevent forgetting of old knowledge. The progress in CL includes, since a few years, also ASR \cite{lifelongasr, eeckt2021continual, eeckt_adapters, sustek22_interspeech, updating_only, other_ucl, ogem, weight_averaging, vandereeckt_interspeech2023}. 

The more challenging scenario within CL is Online Continual Learning (OCL). Unlike offline CL, OCL involves: (i) unknown task boundaries; (ii) a continuous stream of batches from unidentified tasks; and (iii) a single pass over each batch, i.e. accesss to a batch is lost after learning from it. In contrast to offline CL’s sequential task presentation, OCL assumes a non-i.i.d. stream of batches, whose task is unknown to the model, that must be utilized without impairing previous task performance. OCL for ASR remains relatively unexplored, with only a handful of studies addressing it \cite{other_ucl, ogem, vandereeckt_interspeech2023}.

This paper extends the realm of OCL into unsupervised OCL (UOCL) for ASR. UOCL challenges the model to learn from unlabeled batches from new (and unknown) tasks while preserving performance on previously encountered domains. We adopt self-training, where the model generates its own pseudo-labels, to facilitate learning from unlabeled data. Our contributions are as follows: (1) we delve into the problem of UOCL, with a focus on the interplay between forgetting and self-training; (2) we compare various OCL and self-training methods; and (3) we propose a new UOCL method, building upon the work in \cite{vandereeckt_interspeech2023}. Our findings reveal that UOCL experiences significantly less forgetting compared to supervised OCL, often resulting in UOCL methods closely matching the performance of supervised OCL approaches due to the balance between reduced forgetting in UOCL and the effectiveness of learning from ground truth vs. pseudo-labels in supervised OCL. Our proposed extension enhances UOCL performance further, demonstrating a significant reduction in the performance gap with its supervised counterparts, nearly equaling performance of supervised scenarios. This advancement marks a considerable step towards General CL \cite{dark_er}, aiming to enable ASR models to leverage all available data to continually improve without forgetting, resulting in highly robust and versatile models.

\section{Framework \& Prior Research}

\subsection{Problem Statement}

Online CL begins with a model defined by parameters $\theta_0$, trained on a labeled dataset $\mathcal{D}_0$ for task (e.g. certain accent, dialect, text domain, etc.) $\mathcal{T}_0$. Upon deployment, the model processes a continuous, non-i.i.d. stream of data batches $\{\mathcal{B}_i\}_{i>0}$, each related to a specific task $\mathcal{T}_{t_i}$, but without knowledge of which task a batch belongs to. As the model processes a new batch $\mathcal{B}_{i+1}$ with parameters $\theta_i$, it seeks to learn as much as possible, updating to $\theta_{i+1}$, while ensuring it does not forget its knowledge from $\mathcal{T}_0$ or previous batches $\{\mathcal{B}_j\}_{j \in (0, i]}$. Access to the initial data, $\mathcal{D}_0$, and earlier batches is no longer available.

In contrast to earlier OCL research in ASR, which primarily considers batches of labeled data pairs $(X, y)$ \cite{ogem, vandereeckt_interspeech2023}, this paper, similar to \cite{other_ucl}, explores unsupervised OCL, moving beyond the need for labeled data and focusing on scenarios with only unlabeled speech $X$. Our objective is for the model to effectively use unlabeled data, enhancing its performance on all tasks encountered $\mathcal{T}_k$ for $k \geq 0$, while not degrading (and ideally even improving) on past tasks. To learn from unlabeled data, self-training is considered.

\subsection{Related Work}

While UOCL has been explored within computer vision \cite{ucl_prev1, uocl_prev, ucl_prev}, within the context of ASR, it remains, with the exception of \cite{other_ucl}, a largely unexplored area. UOCL builds upon various domains investigated in ASR.

\noindent \textbf{Online Continual Learning.} Though offline CL has garnered significant interest in ASR \cite{lifelongasr, eeckt_adapters, sustek22_interspeech, updating_only, weight_averaging}, exploration of the more demanding online CL scenario remains limited \cite{other_ucl, ogem, vandereeckt_interspeech2023}. Our contribution builds upon the existing work by further advancing OCL into the realm of unsupervised learning.

\noindent \textbf{Self-Training (ST).} ST \cite{self-training} leverages a model initially trained on labeled data to generate pseudo-labels (PLs) for unlabeled data, facilitating subsequent model updates. These PLs can be produced through various methods such as greedy CTC \cite{likhomanenko21b_interspeech, higuchi21_interspeech}, beam search decoding \cite{chen2020, large_scale_asr_domain_adapt}, or by incorporating an external LM \cite{Xu2020, complementary}. ST thus allows the same or a new model to be updated on the pseudo-labeled data.

\noindent \textbf{Unsupervised Domain Adaptation (UDA).} UDA involves adapting a model from a source domain to a target domain using unlabeled data. In the context of ASR, examples include \cite{cmatch, madi, zhu2022boosting}. Although UDA shares similarities with UOCL, it diverges in several key aspects: it assumes that (a) labeled data from the source domain remains available; (b) the model has access to the full unlabeled target training set; (c) only the performance on the target domain matters.

\noindent \textbf{Test-Time Adaptation (TTA).} TTA occurs when the source data is unavailable while adapting to new domains using (unlabeled) test data, often limited in quantity, such as a batch or a single utterance \cite{lin22b_interspeech}. Works for ASR include \cite{lin22b_interspeech, kim2023sgem}. Despite similarities to UOCL, TTA differs significantly: (a) UOCL involves adapting to multiple tasks or domains without forgetting, unlike TTA, which focuses on immediate adaptation with minimal data; (b) TTA adaptations are performed at test time, often based on limited data, whereas OCL adapts during training, handling more extensive tasks.

\section{Methodology}
UOCL combines Self-Training (ST) for generating pseudo-labels with OCL strategies to counteract forgetting. After outlining the model, alongside the ST and OCL methods pertinent to our investigation, we present tailored enhancements to an established OCL technique, optimizing its efficacy for UOCL.

\subsection{Model}
The model considered in this study is an encoder-decoder end-to-end ASR model, characterized by parameters $\theta \in \mathbb{R}^N$. It takes speech $X$ as input, consisting of $F$ speech frames. The model outputs a probability distribution across $C$ word pieces. Incorporating a hybrid approach \cite{hybrid_ctctransformer}, our model integrates losses from both CTC and a decoder, weighted by $c$ and $1-c$, resp., with $c \in [0, 1]$. The model contains (during training) random effects such as dropout \cite{dropout} and data augmentation.
%For speech data $X$ paired with transcriptions $y$, the loss is computed as follows (with $\L_\text{e2e}$ the cross-entropy loss from the decoder, and $\L_\text{ctc}$ the CTC loss):
%\begin{equation}
%    \L(X, y; \theta) = (1 - c) \L_\text{e2e}(X, y; \theta) + c \L_\text{ctc}(X, y; \theta)
%    \label{eq:loss}
%\end{equation}

\subsection{Self-Training Methods}
\label{subsec:st}
We explore these ST methods to generate pseudo-labels (PLs):
\noindent \textit{(Greedy) CTC.} The decoder is not involved in PL generation, which is fast and only based on acoustic evidence, not on a (strong) internal LM in the decoder. 

\noindent \textit{Hybrid.} Combines decoder and CTC to produce PLs, with same weight $c$ as during training and beam size $b$ as during testing. 

%\noindent \textit{Mixed} In some cases, the decoder might generate PLs that are very far (acoustically) from what was actually being said. This could be due to the decoder's bias (from the old task) or due to repetitions in case the utterances differs much from what the decoder was used to during training on the labeled data. We check if this is the case by considering the log probability of the CTC on the Hybrid PLs. If this exceeds a certain threshold $l_\text{CTC}$, we use the CTC PL instead of the Hybrid PL. 

\noindent \textit{BERT-CTC (BCTC).} BERT-CTC \cite{bert_ctc} integrates BERT's \cite{devlin-etal-2019-bert} linguistic knowledge into non-autoregressive CTC predictions for ASR. This approach mitigates the conditional independence assumption by leveraging external LM knowledge, enhancing adaptability across varying text domains without biasing towards the initial task's domain. This ST method should be particularly suited for experiments with text domain shifts. We train the BERT-CTC on top of $\theta_0$ (which we freeze) on task $\T_0$. During UOCL, to update model $\theta_i$ on batch $\B_{i+1}$, we add BERT-CTC (now frozen) on top of $\theta_i$ to generate PLs. 

An external LM trained on new tasks' text domains is not explored, assuming these text domains are unknown a priori. When generating PLs, the model's random effects are disabled.

\subsection{Online CL Methods}
We consider the following OCL methods to apply to UOCL: 

\noindent \textit{Fine-Tuning (FT)}. The model is naively adapted to the new tasks without mechanism to alleviate forgetting. Consequently, Fine-Tuning is expected to suffer greatly from forgetting.

%\noindent \textit{Elastic Weight Consolidation (EWC)} \cite{ewc}. EWC uses the diagonal of the fisher information matrix to estimate the importance of each parameter to old tasks; through a weighted L2 regularization loss the model is then stimulated to learn new tasks by adapting mainly the remaining unimportant parameters (i.e. those with low importance weights). EWC has been extended for OCL as well \cite{pandc}.

%\noindent \textit{Update-Only-Encoders (UOE)} \cite{updating_only}. UOE attempts to overcome forgetting by only adapting the encoder on new tasks, thus freezing the decoder after having trained on the initial task.  

\noindent \textit{Experience Replay (ER)} \cite{er}. ER alleviates forgetting by relying on a small memory set of samples of old tasks, from which it samples a mini-batch to train jointly with the current batch; i.e. samples from old tasks are used to regularize updates to the model. We consider the implementation of \cite{eeckt2021continual} with reservoir sampling \cite{reservoir_sampling} to fill the memory with a size of $M$ utterances. For unlabeled data, the speech and PLs are stored in the memory set. Despite its effectiveness, ER's requirement for data storage can be a major drawback when data storage is restricted.

\noindent \textit{Averaging for Online CL of ASR (AOS)} \cite{vandereeckt_interspeech2023}. AOS employs a dual-model approach: a continually updated adapted model and a stable final model. The adapted model learns from new data batches, with its updates incrementally averaged into the final model. Specifically, after updating the adapted model $\tilde{\theta}_i$ with batch $\B_{i+1}$ to $\tilde{\theta}_{i+1}$, the final model $\theta_i$ updates to $\theta_{i+1} = (1-\eta_{i+1})\theta_i + \eta_{i+1}\tilde{\theta}_{i+1}$. The weight $\eta_{i+1}$ for the encoder is based on number of frames in current batch $F_{i+1}$, total seen frames $F_{1:i}$, and hyper-parameter $\tau$ as $\eta_{\text{enc}, i+1}=\tau F_i / (F_{1:i}+ \tau F_i)$; similar for the decoder, based instead on number of output tokens and using hyper-parameter $\tau_2$. $(\tau, \tau_2)$ are used to control model plasticity. Finally, a knowledge distillation (KD) from final to adapted model is used to regularize the latter. For UOCL adaptation, we thus let the final model generate PLs for the adapted model's training, replacing the KD regularization previously used to temper the adapted model's updates.

\subsection{Our Proposed UOCL Extensions}
\label{subsec:our}
\begin{table*}
    \centering
    \begin{threeparttable}
    \caption{Results of the experiments. Tasks are learned from left to right, with WERs obtained after learning all tasks. \textbf{S}, \textbf{U} and \xmark  in first column refer to, resp., supervised OCL, UOCL and no adaptation. Best UOCL result per task in bold. }
    \begin{tabular}{c c l l c@{\hspace{5pt}} c@{\hspace{5pt}} c@{\hspace{5pt}} c@{\hspace{5pt}} c@{\hspace{5pt}} c@{\hspace{5pt}} c@{\hspace{5pt}} r c@{\hspace{5.6pt}} c@{\hspace{5.6pt}} c@{\hspace{5.6pt}} r}
    \toprule
    %\multirow{3}{*}{\textbf{OCL}} & & \multirow{3}{*}{\textbf{Method}} & \multirow{3}{*}{\textbf{ST}} 
    & &  & & \multicolumn{8}{c}{\textbf{Exp. 1: LIB $\rightarrow$ LIB-APT}} &  \multicolumn{4}{c}{\textbf{Exp. 2: LIB-TED-CV}} \\
    \cmidrule(lr){5-12} \cmidrule(lr){13-16} 
    \textbf{OCL} & & \textbf{Method} & \textbf{ST} & \multicolumn{7}{c}{\textbf{WER per task}} & \multicolumn{1}{c}{\textbf{Avg.}} & \multicolumn{3}{c}{\textbf{WER per task}} & \multicolumn{1}{c}{\textbf{Avg.}}  \\
    \cmidrule(lr){5-11} \cmidrule(lr){12-12} \cmidrule(lr){13-15} \cmidrule(lr){16-16}
     &  & &  & \textbf{LIB} & \textbf{GB/M} & \textbf{US/U} & \textbf{IN/U} & \textbf{IN/M} & \textbf{US/M} & \textbf{GB/U} & \textbf{WER} & \textbf{LIB} & \textbf{TED} & \textbf{CV} & \textbf{WER} \\
    \toprule
    \xmark & 0 & \multicolumn{2}{l}{Initial model $\theta_0$} & 6.8 & 18.4 & 8.2 & 20.0 & 31.7 & 10.9 & 9.5 & 13.18  & 6.4 & 14.0 & 18.8 & 13.08  \\
    \midrule
    \multirow{3}{*}{\textbf{S}} & 1 & \multicolumn{2}{l}{FT} & 8.1 & 11.5 & 8.7 & 15.6 & 23.0 & 11.1 & 6.4 & 10.55\tnote{g} & 8.4 & 12.5 & 16.0 & 12.33\tnote{g}  \\
    & 2 & \multicolumn{2}{l}{ER [$M$=2k]} & 7.1 & 10.4 & 7.7 & \textbf{12.7} & 17.9 & 9.8 & 6.7 & 9.05\tnote{g} & 7.6 & 12.1 & 16.0 & 11.88\tnote{g}  \\
    & 3 & \multicolumn{2}{l}{AOS} & 7.2 & 11.7 & 7.5 & 13.1 & 18.9 & 9.5 & 7.4 & 9.41\tnote{g}  & 6.7 & 12.5 & 16.1 & 11.75\tnote{g} \\
    \midrule 
    \multirow{12}{*}{\textbf{U}} & 4 & \multirow{3}{*}{FT} & CTC & 7.9 & 12.1 & 8.5 & 16.4 & 23.5 & 10.8 & 6.9 & 10.76\tnote{a} & 7.6 & 12.9 & 17.1 & 12.54\tnote{a}  \\
    & 5 & & Hybr. & 8.1 & 11.7 & 8.6 & 16.1 & 23.2 & 11.1 & \textbf{6.9} & 10.72\tnote{a} & 7.6 & 13.4 & 17.1 & 12.70\tnote{a} \\
    %& 6 & & Mixed & 7.7 & 12.0 & 8.3 & 16.0 & 23.5 & 10.8 & \textbf{6.7} & 10.64\tnote{a}  \\
    & 6 & & BCTC & \multicolumn{8}{c}{\textit{Excluded due to lack of text-domain shift}}& 7.7 & \textbf{12.7} & 17.0 & 12.44\tnote{b} \\
    \cmidrule(lr){2-16}
    & 7 & \multirow{2}{*}{ER [$M$=2k]} & CTC & \textbf{7.1} & {11.6} & 7.7 & 14.1 & 19.6 & 9.8 & 7.3 & 9.64 & 6.9 & 12.9 & 17.0 & 12.28 \\
    & 8 & & Hybr. & 7.2 & \textbf{11.2} & 7.8 & 14.1 & \textbf{19.0} & 9.9 & 7.3 & 9.56\tnote{b} & 6.8 & 12.8 & 17.0 & 12.21\tnote{c} \\
    \cmidrule(lr){2-16}
    & 9 & \multirow{2}{*}{AOS} & CTC & 7.2 & 13.1 & \textbf{7.5} & 14.8 & 21.5 & 9.5 & 7.8 & 10.17\tnote{a}  &  6.6 & 13.2 & 17.2 & 12.33  \\
    & 10 & & Hybr. & 7.2 & 12.9 & \textbf{7.5} & 14.9 & 21.6 & 9.6 & 7.8 & 10.17\tnote{a} & 6.8 & 13.7 & 19.5 & 13.32 \\
    \cmidrule(lr){2-16}
    & 11 & \multirow{3}{*}{AOS-U} & CTC & {7.2} & 11.9 & 7.6 & \textbf{13.6} & 19.1 & \textbf{9.3} & 7.5 & \textbf{9.52}\tnote{a,d} & 6.6 & 12.9 & 17.0 & 12.15\tnote{a,e}  \\
    & 12 & & Hybr. & 7.2 & 12.0 & \textbf{7.5} & 13.7 & 19.3 & \textbf{9.3} & 7.5 & 9.55\tnote{a,d} & 6.6 & 13.5 & 17.0 & 12.36\tnote{a} \\
    %& 14 & & Mixed & \textbf{7.1} & 11.7 & 7.6 & \textbf{13.6} & 19.2 & 9.4 & 7.6 & 9.53\tnote{a,d} \\
    & 13 & & BCTC & \multicolumn{8}{c}{\textit{Excluded due to lack of text-domain shift}} & \textbf{6.5} & 12.8 & \textbf{16.8} & \textbf{12.01}\tnote{b,f} \\
    \bottomrule
    \end{tabular}
    \begin{tablenotes}
    \footnotesize
    \item[a] Variations in performance across given ST methods for given UOCL method are not statistically significant.
    \item[b, c] Significantly outperforms other ST methods for given UOCL method with level *** for \textit{b} and * for \textit{c}. 
    \item[d, e] Significantly outperforms all \textit{other} UOCL methods (with any ST) with level *** (\textit{d}) / at least level ** (\textit{e}), except ER + Hybr ($ns$).
    \item[f] Significantly outperforms all combinations of UOCL and ST methods with level ***. 
    \item[g] Significantly outperforms all UOCL variations of given (supervised) OCL method with level ***.
    \end{tablenotes}
    \label{tab:exp}
    \end{threeparttable}
\end{table*}

Building on AOS's proven effectiveness in bypassing the need for past data storage \cite{vandereeckt_interspeech2023}, we adapt it for UOCL, creating AOS-U (AOS-Unsupervised), with simple yet effective modifications:
\begin{enumerate}[label=(\arabic*), leftmargin=*]
    \item The adapted model is our default choice for generating PLs, as it learns quickly, enhancing the quality of PLs rapidly.
    \item Despite utilizing the adapted model for PL generation, we still omit the  KD regularization to avoid impeding its learning progress, ensuring optimal PL performance.
    \item Before updating, the model processes each batch multiple times ($K>1$), utilizing randomness (dropout, data augmentation) to enhance adaptation from unlabeled data. This approach, while still updating the model only once per batch, enriches learning by presenting varied data representations. We take $K=2$.
\end{enumerate}

\section{Experiments}
All experiments were done in ESPnet2 \cite{watanabe2018espnet}. For 
all detailed information, we refer to our Github repository 
\footnote{{https://github.com/StevenVdEeckt/unsupervised-ocl-for-asr}}.
%\footnote{\href{https://github.com/StevenVdEeckt/unsupervised-ocl-for-asr}{https://github.com/StevenVdEeckt/unsupervised-ocl-for-asr}}.

\noindent \textbf{Data.} We consider two experiments: (\textit{Exp. 1}) Adaptation from Librispeech-360h (LIB) \cite{librispeech} to six Libri-Adapt (LIB-APT) \cite{libri_adapt} tasks (9k-30k utterances each), showcasing variations in microphone (USB [U], Matrix [M]) and accent (US [US], Indian [IN], British [GB] English); (\textit{Exp. 2}) LIB to TEDLIUM-3 (TED) \cite{tedlium} to Common Voice-English (CV) \cite{commonvoice}, learned in this order, considering a subset from TED (100k utterances) and three accents from CV (135k utterances) -- this path underscores a text domain shift, marked by CV's distinct shorter sentences with large unique vocabulary, and accent shifts, plus TED's more spontaneous speech.

%\noindent \textbf{Data.} We consider following experiments:
%\begin{enumerate}
%    \item Exp. 1: Adaptation from Librispeech-360h (LIB) to six Libri-Adapt (LIB-APT) \cite{libri_adapt} tasks (9k-30k utterances each), showcasing variations in microphone (USB [U], Matrix [M]) and accent (US [US], Indian [IN], British [GB] English);
%    \item Exp. 2: LIB to TEDLIUM-3 (TED) \cite{tedlium} to Common Voice (CV) \cite{commonvoice}, learned in this order, considering a subset from TED (100k utterances) and three accents from CV (135k utterances). This path underscores a text domain shift, marked by CV's distinct shorter sentences with large unique vocabulary, and accent shifts, plus TED's more spontaneous speech.
%\end{enumerate}
%Batches within each task are sorted by speaker (both dimensions unknown to the model), heightening the challenge and potential for intra-task forgetting. Evaluation employs the LIB test-clean set for Exp. 1, and test-other for Exp.2.

\noindent \textbf{Model.} We consider two models. For Exp. 1, starting from MFCC features, we use 12 Conformer \cite{conformers} encoder and 6 Transformer \cite{transformer} decoder layers, each with dimension 2048 and 4 attention heads with dimension 256. For Exp. 2, we start from pre-trained HuBERT \cite{hubert} features, using HuBERT-Large trained on Libri-Light \cite{librilight} from \cite{hubert} as frontend of a model with 6 Conformer encoder and 6 Transformer decoder layers. A linear pre-encoder is added to map the HuBERT features to the same dimension (80) as the MFCC features. The HuBERT model is frozen, while the linear pre-encoder is only trained on the initial task $\T_0$. For both models,  the weight of CTC is $c=0.3$, we use SentencePiece \cite{sentencepiece} to generate $C=5000$ output tokens on initial task $\T_0$ and SpecAugment \cite{specaugment} is used as data augmentation while dropout rate is set to $p=0.1$. Both models (with, resp., $N=47\text{M}$ and $N=349\text{M}$ [of which $32\text{M}$ trainable] parameters) are trained for 80 epochs on $\T_0$. For OCL, batch size is $10$ and $20$ for Exp. 1 and Exp. 2, resp., and a SGD optimizer with learning rate $0.01$ is used. To evaluate the model, we use the same weight $c=0.3$ for CTC in addition to beam search decoding with beam size $b=1$. For BERT-CTC, we set number of iterations to $5$. We consider BERT-CTC only for Exp. 2, in which there is a text domain shift.

\noindent \textbf{Hyper-parameters.} We optimized each method's hyper-parameters by identifying their best values on a preliminary 'test' experiment, involving LIB-APT and CV tasks (for Exp. 1 and Exp. 2, resp.) not featured in our main experiments.

\noindent \textbf{Metrics.} We report WER (in $\%$, evaluated after processing all batches) per task and averaged over all tasks. Using Wilcoxon signed-rank test on the number of errors per utterance \cite{Strik2000ComparingTR}, we perform significance testing on average WER, considering significance levels $\alpha=0.05$ (*), $\alpha=0.01$ (**) and $\alpha=0.001$ (***), or $ns$ for non-significant ($\alpha > 0.05$).

\section{Results}

Table \ref{tab:exp} presents the results of the two experiments.

\subsection{Exp. 1: LIB to LIB-APT}

\noindent \textbf{Supervised methods.} Compared to the \textit{Initial model}, all supervised methods enhance task performance but exhibit varying degrees of forgetting. ER, benefiting from the storage of past data, achieves the best performance, while AOS, though worse than ER, significantly improves upon the naive FT.

\noindent \textbf{Improvement on all tasks with CTC PL.} Without labels, the UOCL methods manage to enhance performance across all tasks relative to the \textit{Initial model}, including on \textit{IN/M} with a high initial WER. This improvement persists despite the model encountering each batch only once and generating its own PLs. 

\noindent\textbf{Less forgetting for UOCL with FT.} In the UOCL setting, FT using CTC PL exhibits reduced forgetting compared to supervised FT, surpassing it on LIB, US/U, and US/M (for which supervised FT forgets more) and almost matching its average WER. This suggests that for FT, less forgetting from self-training in UOCL nearly counterbalance the superior learning from ground truth labels in supervised OCL. This is not the case for ER and AOS, which mitigate forgetting and for which the absence of ground truths is thus a larger disadvantage; hence the larger gap with their supervised counterpart.

\noindent \textbf{CTC vs. Hybrid PL.} Using Hybrid PL, which includes an internal LM tailored to the new tasks' text domain (the same as $\T_0$), does not significantly improve performance compared to the simpler and faster CTC PL. As its results on LIB, US/U, and US/M illustrate, Hybrid PL tends to forget more than CTC PL. This likely happens because the encoder plays a crucial role in the UOCL process, both in learning new tasks and in reducing forgetting. The decoder, positioned later in the process, can adjust some CTC predictions with its internal LM, which might help to learn the new tasks but also might cause more forgetting (when these PLs are used in the CTC loss). These effects seem to cancel each other out. To explore this idea, we experimented with FT, allowing CTC and decoder to generate and learn from their own PLs, i.e., the PLs made by the decoder are only used in the decoder loss. This results in average WER of 10.61, a significant improvement over CTC and Hybrid PL for FT, combining the advantages from both. Its WERs on LIB, US/U and US/M, are, resp., 7.9, 8.4 and 10.6, illustrating less forgetting. Note, finally, that ER is the only UOCL method whose Hybrid PL significantly outperforms CTC PL. Thanks to its access to past (labeled) data, ER is able to profit from the advantages of Hybrid PL while mitigating the disadvantages.

\noindent \textbf{AOS-U outperforms AOS.} Our enhancements to AOS for the UOCL scenario (AOS-U) significantly boost unsupervised AOS's effectiveness, narrowing the performance difference with its supervised counterpart. AOS-U shows improvements in five of the seven tasks, excelling in tasks with initially high error rates (IN/U and IN/M). Although there was a notable performance gap between ER and AOS in both supervised and unsupervised settings, AOS-U matches ER's unsupervised performance without the need to store past data. An ablation study in Table \ref{tab:ablation} reveals AOS-U's strengths come from both applying the adapted model for pseudo-labeling and processing each batch through the model multiple times.

\subsection{Exp. 2: LIB to CV}

\noindent \textbf{Less forgetting for UOCL.} Unlike in the previous experiment, when adapting LIB to TED and CV, there exist a text domain shift, which causes forgetting in supervised OCL for all methods, though to different extents. When instead learning from PLs (using CTC PLs) in UOCL, we observe a significant reduction in forgetting, not only for the naive FT but also for ER and AOS(-U). For FT, the disadvantage of learning from noisy PLs and the advantage of resulting reduced forgetting almost balance each other out, comparing its performance for UOCL vs. supervised OCL. As in Exp. 1, for ER and AOS(-U), since they mitigate forgetting, the aforementioned disadvantage outweighs the advantage, and, consequently, the gap between their UOCL and supervised OCL performance is larger.

\noindent \textbf{CTC vs. Hybrid ST Method.} Even more than in the previous experiment, we observe that, though Hybrid PLs are on average better than CTC PLs, Hybrid ST method suffers from problems which cause its performance to be worse, not better, than CTC. In this experiment, we observe that Hybrid PL, in exceptional cases (1-5$\%$ of utterances), tends to generate PLs with very little acoustic evidence, due to e.g. repetitions. This is the case when adapting to both TED (for which utterances are sometimes cut-off sentences) and CV (for which utterances are much shorter). CTC PL thus seems to be more robust for UOCL, even if on average its PLs have higher WER than Hybrid PL. Its advantage is that its predictions are based on direct acoustic evidence, which also results in less forgetting in UOCL.

\noindent \textbf{AOS-U outperforms AOS for UOCL.} As in the previous experiment, AOS-U significantly outperforms AOS, though in this case, for CTC PL, it is only the usage of the adapted model as generator of PLs that is significant, as illustrated by Table \ref{tab:ablation}. 

\noindent \textbf{BERT-CTC PL outperforms other ST Methods.} As explained in Sec. \ref{subsec:st}, BERT-CTC, starting from the CTC prediction, allows the model to rely on external LM information without having to make assumptions about the text domain. While the decoder, which relies on an internal LM, has a bias towards the old text domain -- no longer valid for the new tasks -- BERT-CTC suffers much less from  biases. While BERT-CTC decoding using \textit{Initial model} $\theta_0$ results in higher WER on the initial task $\T_0$ compared to Hybrid decoding (4.9 vs. 5.4 on LIB's \textit{dev-other}), its WER on the unseen tasks is better. Consequently, it generates better PLs which result in better overall performance. Compared to Hybrid PL, BERT-CTC PL starts directly from CTC and encoder features, which can attend to BERT features, generating improvements over CTC PL for which there exists more acoustic evidence compared to Hybrid PL. Increased forgetting due to using external information from BERT is minimal, as shown by FT (AOS-U even improves on $\T_0$). Overall, thus, AOS-U with BERT-CTC PLs outperforms all other UOCL methods, and almost matches the performance of supervised ER, which, unlike AOS-U, can take advantage of both some stored past data and ground truth labels, and AOS. 

\begin{table}
    \centering
    \caption{Ablation study assessing impact of enhancements detailed in Sec. \ref{subsec:our} for Experiments 1 and 2. }
    \begin{tabular}{l l l}
    \toprule
    \multirow{2}{*}{\textbf{Method}} & \multicolumn{2}{c}{\textbf{Average WER}} \\
    \cmidrule(lr){2-3}
    & \multicolumn{1}{l}{\textbf{Exp. 1}} & \multicolumn{1}{l}{\textbf{Exp. 2}} \\
    \toprule
    AOS w. CTC-PL & 10.17 & 12.33 \\
    + (1) + (2) & \phantom{1}9.70*** &  12.17***  \\
    + (1) + (2) + (3) & \phantom{1}9.52*** &  $12.15^{\text{ns}}$\phantom{*}  \\
    + (1) + (3) & \phantom{1}$9.49^\text{ns}$ &  $12.20^{\text{ns}}$\phantom{*}  \\
    \bottomrule
    \end{tabular}
    \label{tab:ablation}
\end{table}

\section{Conclusion}

We study and advance unsupervised OCL (UOCL) for ASR, enabling learning from non-i.i.d. streams of unlabeled data with minimal forgetting. Our exploration into UOCL reveals that it substantially mitigates forgetting, enabling models to closely align with supervised performance despite solely relying on unlabeled data. By implementing self-training techniques, we demonstrate the potential for ASR systems to continually adapt and learn from new, unlabeled tasks while preserving knowledge across old domains. Our findings highlight the robustness of CTC-based pseudo-labeling, focusing on acoustic evidence for consistent performance, and the BERT-CTC approach as an effective solution for adapting to text-domain shifts. Our proposed UOCL extensions outperform existing approaches, offering a robust solution for ASR systems in label-scarce environments. Our work paves the way for more resilient and adaptable ASR systems, capable of evolving in dynamic environments without the need for constant retraining with labeled data.

\section{Acknowledgements}
Research supported by Research Foundation Flanders (FWO) under grant S004923N of the SBO programme.

\bibliographystyle{IEEEtran}
\bibliography{mybib}

\end{document}